\documentclass[twoside,twocolumn,9pt]{article}
\usepackage{extsizes}
\usepackage[super,sort&compress,comma]{natbib} 
\usepackage[version=3]{mhchem}
\usepackage[left=1.5cm, right=1.5cm, top=1.785cm, bottom=2.0cm]{geometry}
\usepackage{balance}
\usepackage{mathptmx}
\usepackage{sectsty}
\usepackage{graphicx}
\usepackage{lastpage}
\usepackage{booktabs}
\usepackage{subfig}
\usepackage{subcaption}
\usepackage[format=plain,justification=justified,singlelinecheck=false,font={stretch=1.125,small,sf},labelfont=bf,labelsep=space]{caption}
\usepackage{float}
\usepackage{fancyhdr}
\usepackage{fnpos}
\usepackage{lipsum}
\usepackage[english]{babel}
\addto{\captionsenglish}{%
  
}
\usepackage{array}
\usepackage{droidsans}
\usepackage{charter}
\usepackage[T1]{fontenc}
\usepackage[usenames,dvipsnames]{xcolor}
\usepackage{setspace}
\usepackage{bm}
\usepackage[compact]{titlesec}
\usepackage[hidelinks]{hyperref}

\usepackage{epstopdf}

\definecolor{cream}{RGB}{222,217,201}
\definecolor{delftblue}{RGB}{31, 48, 94}
\definecolor{cafenoir}{RGB}{75, 54, 33}
\definecolor{jet}{RGB}{52, 52, 52}
\definecolor{col1}{rgb}{0,0.466667,0.733333}
\definecolor{col2}{rgb}{0.2,0.733333,0.933333}
\definecolor{col3}{rgb}{0,0.6,0.533333}
\definecolor{col4}{rgb}{0.933333,0.466667,0.2}
\definecolor{col5}{rgb}{0.8,0.2,0.0666667}
\definecolor{col6}{rgb}{0.933333,0.2,0.466667}
\definecolor{col7}{rgb}{0.733333,0.733333,0.733333}
\definecolor{col8}{rgb}{0.392157,0.392157,0.392157}

\begin{document}

\pagestyle{fancy}
\thispagestyle{plain}
\fancypagestyle{plain}{
\renewcommand{\headrulewidth}{0pt}
}

\makeFNbottom
\makeatletter
\renewcommand\LARGE{\@setfontsize\LARGE{15pt}{17}}
\renewcommand\Large{\@setfontsize\Large{12pt}{14}}
\renewcommand\large{\@setfontsize\large{10pt}{12}}
\renewcommand\footnotesize{\@setfontsize\footnotesize{7pt}{10}}
\makeatother

\renewcommand{\thefootnote}{\fnsymbol{footnote}}
\renewcommand\footnoterule{\vspace*{1pt}%
\color{cream}\hrule width 3.5in height 0.4pt \color{black}\vspace*{5pt}} 
\setcounter{secnumdepth}{5}

\makeatletter 
\renewcommand\@biblabel[1]{#1}            
\renewcommand\@makefntext[1]%
{\noindent\makebox[0pt][r]{\@thefnmark\,}#1}
\makeatother 
\renewcommand{\figurename}{\small{Fig.}~}
\sectionfont{\sffamily\Large}
\subsectionfont{\normalsize}
\subsubsectionfont{\bf}
\setstretch{1.125} 
\setlength{\skip\footins}{0.8cm}
\setlength{\footnotesep}{0.25cm}
\setlength{\jot}{10pt}
\titlespacing*{\section}{0pt}{4pt}{4pt}
\titlespacing*{\subsection}{0pt}{15pt}{1pt}

\fancyfoot{}
\fancyfoot[LO,RE]{\vspace{-7.1pt}\includegraphics[height=9pt]{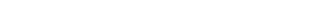}}
\fancyfoot[CO]{\vspace{-7.1pt}\hspace{11.9cm}\includegraphics{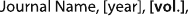}}
\fancyfoot[CE]{\vspace{-7.2pt}\hspace{-13.2cm}\includegraphics{head_foot/RF}}
\fancyfoot[RO]{\footnotesize{\sffamily{1--\pageref{LastPage} ~\textbar  \hspace{2pt}\thepage}}}
\fancyfoot[LE]{\footnotesize{\sffamily{\thepage~\textbar\hspace{4.65cm} 1--\pageref{LastPage}}}}
\fancyhead{}
\renewcommand{\headrulewidth}{0pt} 
\renewcommand{\footrulewidth}{0pt}
\setlength{\arrayrulewidth}{1pt}
\setlength{\columnsep}{6.5mm}
\setlength\bibsep{1pt}

\makeatletter 
\newlength{\figrulesep} 
\setlength{\figrulesep}{0.5\textfloatsep} 

\newcommand{\topfigrule}{\vspace*{-1pt}%
\noindent{\color{cream}\rule[-\figrulesep]{\columnwidth}{1.5pt}} }

\newcommand{\botfigrule}{\vspace*{-2pt}%
\noindent{\color{cream}\rule[\figrulesep]{\columnwidth}{1.5pt}} }

\newcommand{\dblfigrule}{\vspace*{-1pt}%
\noindent{\color{cream}\rule[-\figrulesep]{\textwidth}{1.5pt}} }

\makeatother

\twocolumn[
  \begin{@twocolumnfalse}
{\includegraphics[height=30pt]{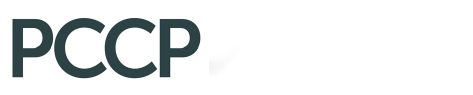}\hfill\raisebox{0pt}[0pt][0pt]{\includegraphics[height=55pt]{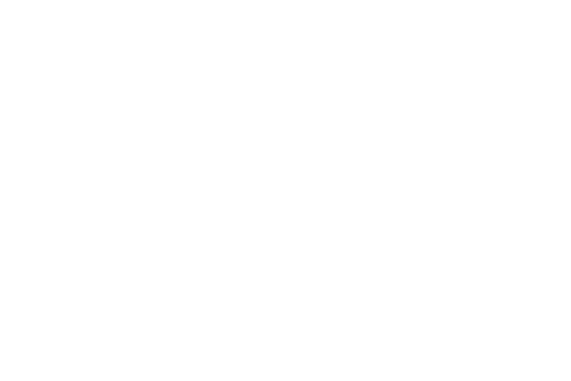}}\\[1ex]
\includegraphics[width=18.5cm]{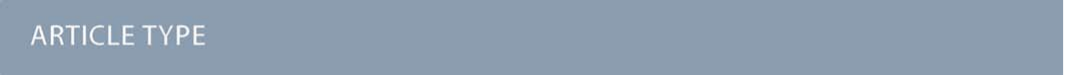}}\par
\vspace{1em}
\sffamily
\begin{tabular}{m{4.5cm} p{13.5cm} }

\includegraphics{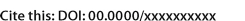} & \noindent\LARGE{\textbf{Triplets in the cradle: ultrafast dynamics in a cyclic disulfide$^\dag$}} \\ 
\vspace{0.3cm} & \vspace{0.3cm} \\

 & \noindent\large{James Merrick\textit{$^{a}$}, Lewis Hutton\textit{$^{a}$}, Joseph C. Cooper\textit{$^{a}$}, Claire Vallance$^{\ast}$\textit{$^{a}$}}, Adam Kirrander$^{\ast}$\textit{$^{a}$} \\

\includegraphics{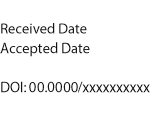} & \noindent\normalsize{The effect of spin-orbit coupling on the \textit{"Newton's cradle"}-type photodynamics in the cyclic disulfide 1,2-dithiane (\ce{C4H8S2}) is investigated theoretically. We consider excitation by a $290$ nm laser pulse and simulate the subsequent ultrafast nonadiabatic dynamics by propagating surface-hopping trajectories using SA(4|4)-CASSCF(6,4)-level electronic structure calculations with a modified ANO-R1 basis set. Two simulations are run: one with singlet states only, and one with both singlet and triplet states. All trajectories are propagated for $1$ ps with a $0.5$ fs timestep. Comparison of the simulations suggests that the presence of triplet states depletes the singlet state population, with the net singlet and triplet populations at long times tending towards their statistical limit. Crucially, the triplet states also hinder the intramolecular thiyl radical recombination pathway {\it{via}} the efficient intersystem crossing between the singlet and triplet state manifolds.} \\
\end{tabular}

 \end{@twocolumnfalse} \vspace{0.6cm}

  ]

\renewcommand*\rmdefault{bch}\normalfont\upshape
\rmfamily
\section*{}
\vspace{-1cm}


\footnotetext{\textit{$^{a}$~ Department of Chemistry, Physical and Theoretical Chemistry Laboratory, University of Oxford, South Parks Road, Oxford, OX1 3QZ, UK. Fax: +44 (0)1865 275400; Tel: +44 (0)1865 275400; E-mail: adam.kirrander@chem.ox.ac.uk}}

\footnotetext{\dag~Electronic Supplementary Information (ESI) available. See DOI: }




\section{Introduction}

Disulfide linkages are a common chemical motif in proteins and are usually formed {\it{via}} covalent coupling of thiol functional groups situated on spatially adjacent cysteine amino acids.\cite{disulphide-review-2002} As a consequence of the  thermal and photolytic stability of cross-linked cysteine thiol groups, and their role as a conformational lock, disulfide linkages are notably important for stabilising tertiary and quaternary protein structures,\cite{COWGILL196737, sevier_formation_2002,qiu_ultrafast_2008} helping to maintain biological function in the native environment.

Conversely, linear alkyl disulfides are unstable with respect to heat and ultraviolet (UV) light both in solution and in the gas phase.\cite{lyons_photolysis_1948,Callear1970,rinker_photodissociation_2005, luo_theoretical_2009,ochmann_uv-photochemistry_2018} Cyclic disulfides, in which the disulfide bond is part of a ring, exhibit greater thermal and photolytic stability in both experimental and computational studies, although initial disulfide cleavage is still observed.\cite{stephansen_photostability_2014, solling_dithiane_2018, stephansen_surprising_2012, cao_disulfide_2019,rankine_theoretical_2016,ibele_comparing_2021,lassmann_extending_2022,middleton_--fly_2023,wang_intrinsic_2023,dithiane_2025_lingyu} This enhanced stability has been attributed to the carbon backbone; following initial sulfur-sulfur photoinduced homolysis, the backbone tethers the two dithiyl radicals so that they may return to sufficient proximity for radical recombination to occur.

\begin{figure}[btp]
\centering
\includegraphics[width=0.45\textwidth]{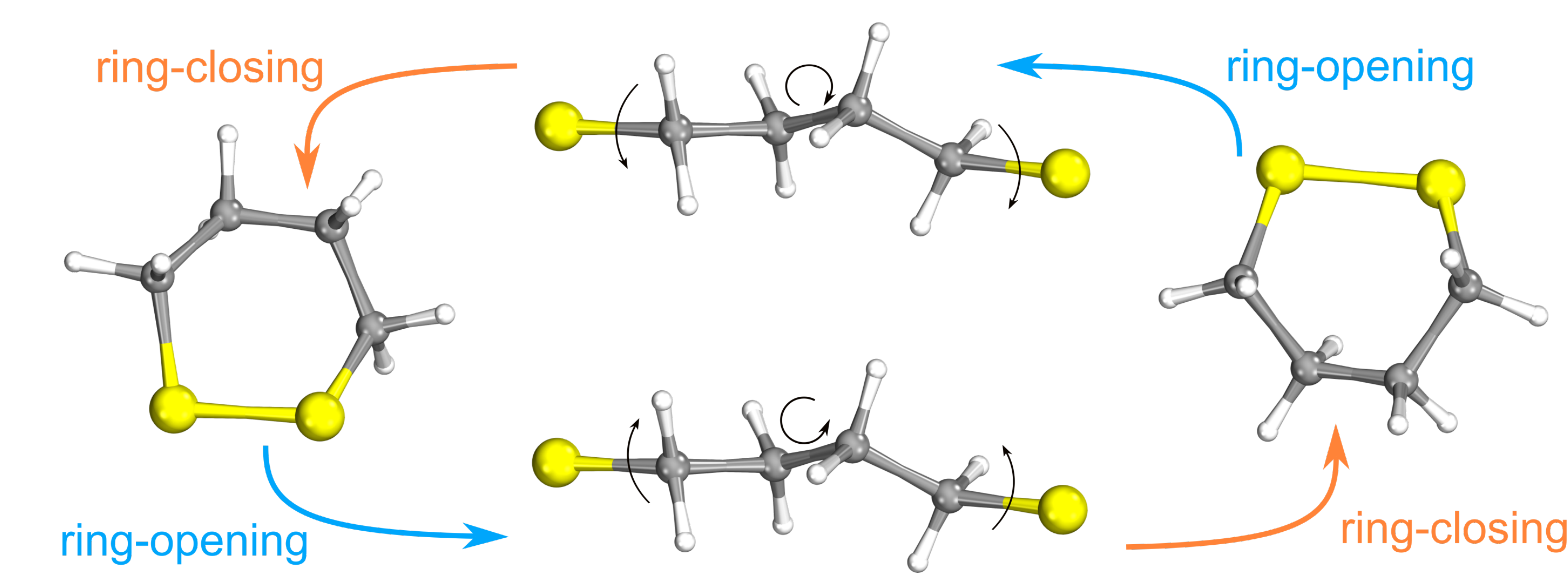}
  \caption{A schematic representation of the \textit{"Newton's cradle"} motion in photoexcited 1,2-dithiane, whereby the molecule repeatedly rotates about the carbon framework to bring the two sulfur atoms back into close proximity before the molecule springs open to reform the linear biradical.}
  \label{fig:schematic}
\end{figure}

The molecule 1,2-dithiane (\ce{C4H8S2}) can be thought of as a six-membered hydrocarbon ring in which two adjacent methylene groups have been replaced by sulfur atoms, as can be seen in Fig.\ \ref{fig:schematic}. It serves as a convenient model for studying the photochemistry of constrained disulfide systems such as those found in proteins or other structurally-constrained environments. Its small size compared to disulfides in proteins -- where polar and charged functional groups, as well as sterically bulky neighbouring functional groups, may affect the \ce{S-S} bond-breaking process -- renders the molecule suitable for accurate computational studies of the sub-picosecond dynamics, and allows the effect of UV light on the \ce{S-S} moiety to be studied in isolation.

\begin{figure*}[h]
\centering
\includegraphics[width=0.89\textwidth]{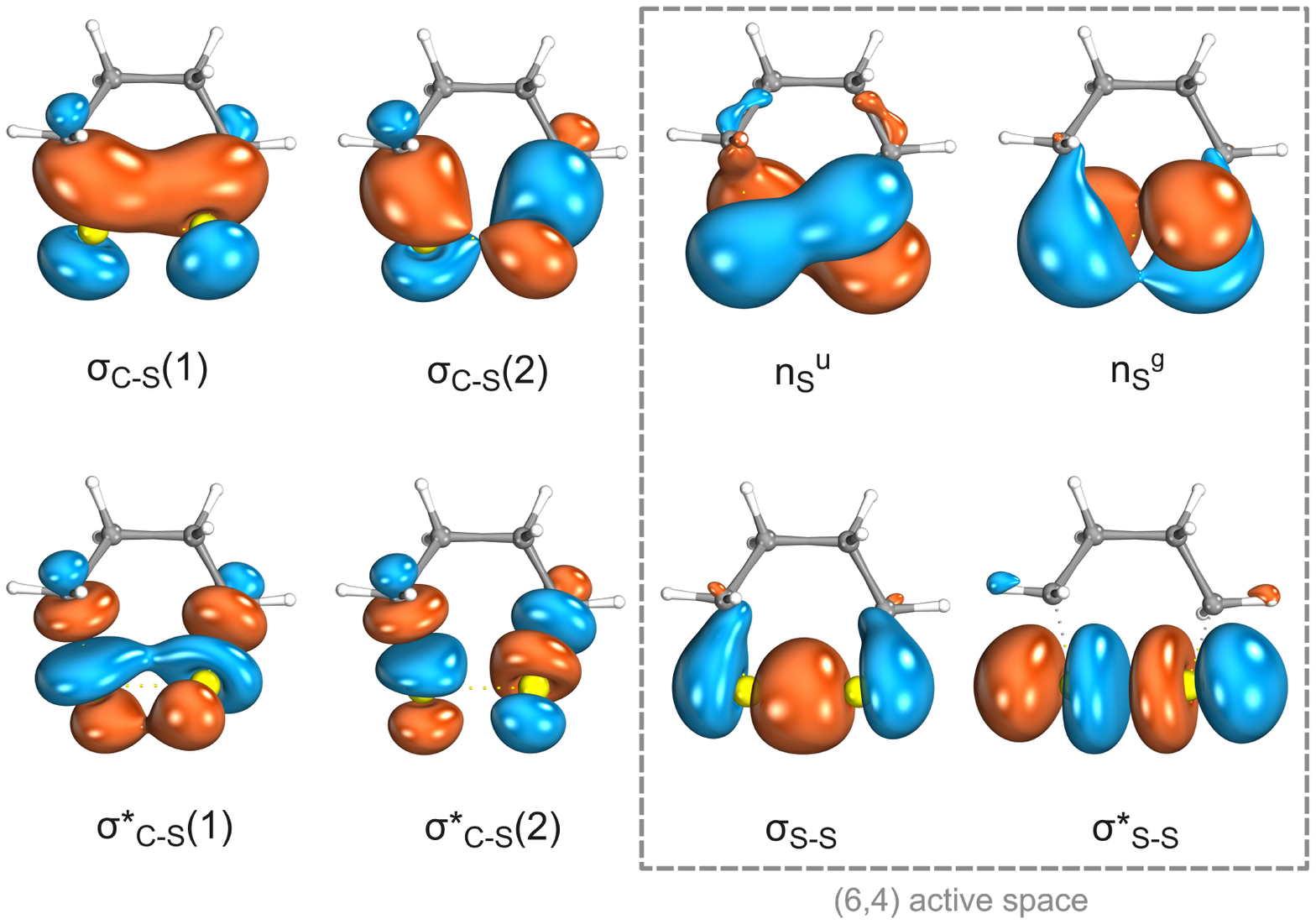}
  \caption{The orbitals used in the active spaces for the electronic structure calculations in 1,2-dithiane. The complete set used in the larger (10,8) active space corresponds to all orbitals shown. The four orbitals included in the smaller (6,4) active space are the subset in the grey-dashed box on the right.}
  \label{fig:orbitals}
\end{figure*}

The photoinduced dynamics of 1,2-dithiane upon UV excitation,
and the effect of molecular geometry on the observed photodynamics of disulfides generally, have both been subject of previous experimental and computational studies.\cite{stephansen_photostability_2014, ibele_comparing_2021, lassmann_extending_2022,stephansen_surprising_2012,rankine_theoretical_2016,middleton_--fly_2023,cao_disulfide_2019,solling_dithiane_2018,wang_intrinsic_2023, dithiane_2025_lingyu} Studies on 1,2-dithiane in particular have found that when the photodynamics is initiated on the first excited singlet electronic state \ce{S_1}, ultrafast \ce{S-S} homolysis occurs on a sub-100 fs timescale and results in ring-opening.\cite{stephansen_surprising_2012,rankine_theoretical_2016,cao_disulfide_2019,ibele_comparing_2021}  The dissociated sulfur termini remain tethered by the carbon backbone and, \textit{via} torsion about the carbon backbone, encounter each other again "on the other side" of the molecule. This results in an oscillatory ring-opening/closing motion with an approximate period of $350$ fs. The characteristic motion, colloquially referred to as a molecular \textit{"Newton's cradle"}, is schematically illustrated in Fig.\ \ref{fig:schematic}. Each time the two sulfur atoms return to each other, radical recombination of the \ce{S-S} bond can occur on the \ce{S_0} ground electronic state. Subsequently, the molecule can spring open again or remain in the vibrationally hot cyclic ground state. The Newton's cradle dynamics is gradually damped by recombination and energy dispersion, the latter becoming intramolecular vibrational energy redistribution (IVR) on longer timescales.\cite{rankine_theoretical_2016}

With one exception,\cite{cao_disulfide_2019} most theoretical studies have considered only singlet excited states in computational nonadiabatic dynamics studies of 1,2-dithiane.\cite{rankine_theoretical_2016,ibele_comparing_2021, lassmann_extending_2022,middleton_--fly_2023,solling_dithiane_2018} However, due to the presence of the sulfur atoms, moderate spin-orbit coupling (SOC)\cite{pauli_zur_1927,soc_review_2012} may be expected to play a role in the photodynamics of 1,2-dithiane. Intersystem crossing (ISC) is known to occur at short timescales in sulfur-containing molecules, allowing ISC to compete with internal conversion (IC).\cite{KirranderMinnsPRL2018,Bellshaw2019} Furthermore, the previous study which accounted for triplet states in simulations of 1,2-dithiane evidences efficient ISC from singlet states to triplet states in the short-time regime -- sub-70 fs -- following photoexcitation.\cite{cao_disulfide_2019} In the present work, the effect of the triplet states on the photodynamics in 1,2-dithiane -- both with respect to electronic state populations and nuclear dynamics -- is considered by comparing otherwise identical 
nonadiabatic dynamics simulations with and without triplet states. The simulations are carried out using Tully's fewest-switches surface-hopping method,\cite{tully_molecular_1990} and both simulations are propagated for $1$ ps to allow both short-time and intermediate-time processes to be examined. Finally, X-ray scattering patterns for the two resulting trajectory ensembles are calculated in order to compare how any differences in nuclear dynamics between the two sets may manifest in ultrafast scattering experiments.\cite{Odate2023review,SimmermacherCh3Kasra,YongCh9Kasra,Weber2024}

\section{Methods}

\subsection{Electronic structure calculations}

The electronic structure calculations are performed at the complete active space self-consistent field (CASSCF) level of theory\cite{roos_complete_1980} using the OpenMolcas v23.02 software package\cite{li_manni_openmolcas_2023} unless otherwise stated. As discussed below, we employ a (6,4) active space for the electronic structure used in the simulations. This active space, shown in Fig.\ \ref{fig:orbitals}, comprises $6$ electrons distributed across $4$ orbitals; the orbitals include \ce{S-S} $\sigma$-bonding and $\sigma^*$-antibonding orbitals ($\sigma_{\mathrm{S-S}}$ and $\sigma^*_{\mathrm{S-S}}$), and two in-phase and out-of-phase combinations of lone-pair orbitals on the adjacent sulfur atoms, labelled n$_\mathrm{S}^\mathrm{u}$ and n$_\mathrm{S}^\mathrm{g}$, respectively. State-averaging is performed over the four lowest energy electronic states in each spin-manifold, \textit{i.e.}\ SA(4) for singlets only, and SA(4|4) for both singlets and triplets. For the electronic structure calculations involving triplets, the SOC Hamiltonian is calculated in the adiabatic-state basis. The spin-orbit states and associated spin-orbit coupling matrix elements (SOCMEs) are then found by diagonalising the total Hamiltonian. This is achieved with the restricted active-space self-interaction (RASSI) program within OpenMolcas, which uses atomic mean-field integrals (AMFIs) within the exact-two-component (X2C) scalar relativistic integrals framework to decouple the total four-component Dirac Hamiltonian.\cite{roos_2004_amfi,hess_1996, peng_exact_2012}

In the simulations, the ANO-R1 basis set is employed.\cite{zobel_ano-r_2020} The basis is truncated by removing the polarisation functions on all hydrogen atoms. This is justified because the hydrogen atoms are not expected to play a significant role in the ring-opening photodynamics of 1,2-dithiane upon photo-excitation at $290$ nm, as supported by previous studies.\cite{ibele_comparing_2021, lassmann_extending_2022,stephansen_surprising_2012,rankine_theoretical_2016,middleton_--fly_2023,cao_disulfide_2019} The truncated ANO-R1 basis is denoted ANO-R1(t) for brevity. The resulting electronic structure method is SA(4|4)-CASSCF(6,4)/ANO-R1(t), and is the method used henceforth unless otherwise indicated. Additional electronic structure benchmarks are provided in the Supplementary Information (SI), in which: (i) calculations using ANO-R1 and ANO-R1(t) are compared against a range of basis sets; and (ii) calculations are compared employing both the (6,4) active space and a larger (10,8) active space in which two sets of \ce{C-S} $\sigma$ and $\sigma^*$ orbitals are included in the active space --- see Fig.\ \ref{fig:orbitals}.

\begin{figure*}[h]
\centering
\includegraphics[trim={0 0 0 0},clip, width=0.98\textwidth]{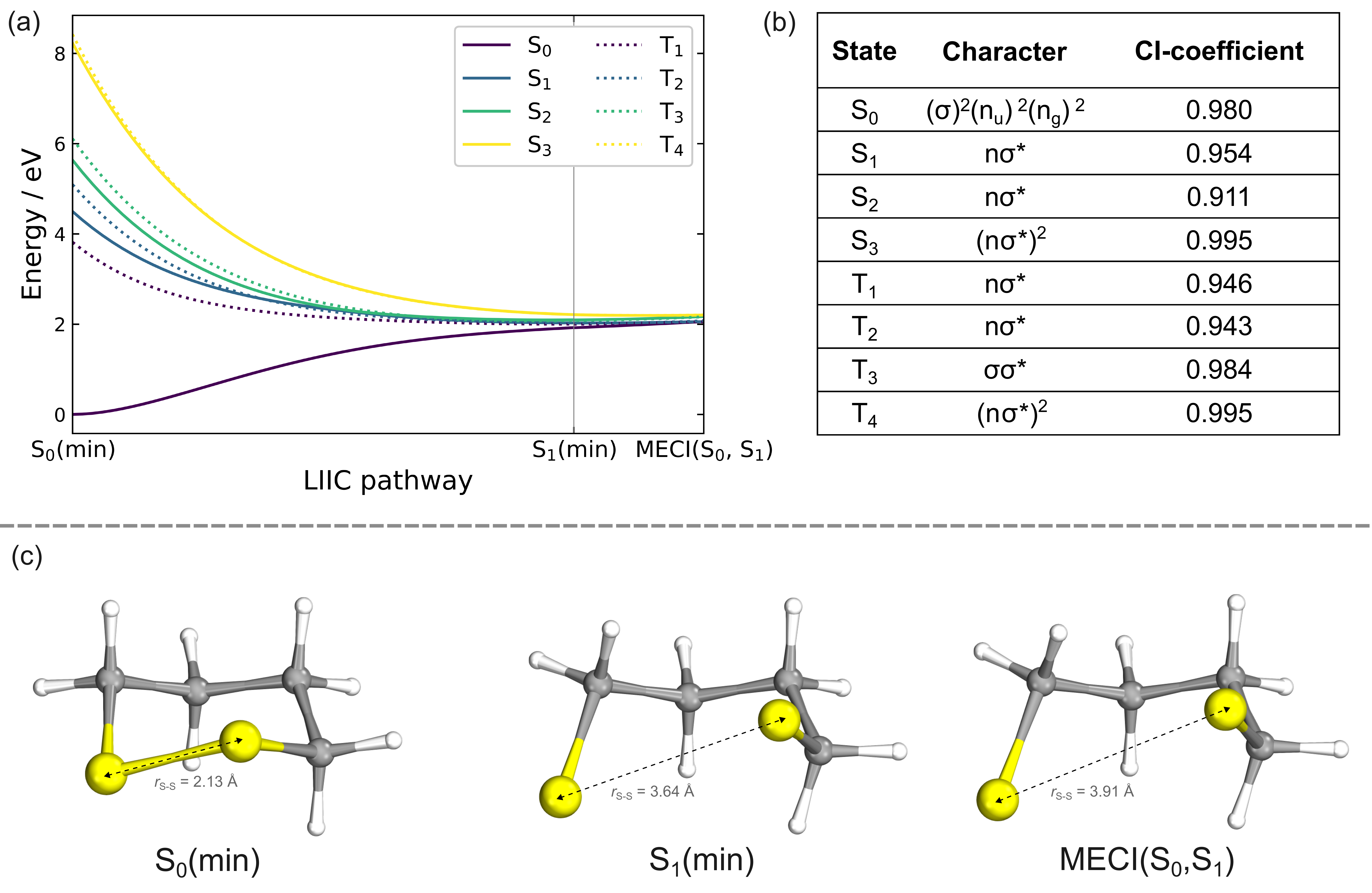}
  \caption{A summary of static results in 1,2-dithiane. (a) Potential energy cuts for the four lowest-energy singlet states \ce{S_0}-\ce{S_3} and triplet states \ce{T_1}-\ce{T_4} along the LIIC pathway connecting the \ce{S_0} minimum S$_{0}$(min) to the \ce{S_1} minimum S$_{1}$(min), and then onwards to the minimum energy conical intersection between the S$_0$ and S$_1$ states, MECI(\ce{S_0},\ce{S_1}). (b) The dominant character and corresponding CI-coefficient for each electronic state at the S$_0$(min) molecular geometry. (c) The optimised molecular geometries S$_{0}$(min) (left), S$_{1}$(min) (centre), and MECI(\ce{S_0},\ce{S_1}) (right).}
  \label{fig:LIIC}
\end{figure*}

Molecular geometries are optimised using SA(4|4)-CASSCF(6,4)/ANO-R1(t). The ground state and the first singlet excited state equilibrium structures, denoted \ce{S_0}(min) and \ce{S_1}(min), are confirmed to correspond to true minima by a frequency calculation (see SI for details). The minimum-energy conical intersection (MECI) between the \ce{S_0} and \ce{S_1} states, denoted MECI(\ce{S_0},\ce{S_1}), is also located. The electronic structure calculations are benchmarked at a range of molecular geometries determined by linear interpolation in internal coordinates (LIIC) between the optimised geometries, as shown in Fig.\ \ref{fig:LIIC}. It is worth noting that the linear interpolation of molecular geometries means that all internal coordinates may change across the LIIC pathway; that is, the LIIC does not strictly correspond to a minimum energy path between the optimised geometries.

Electronic structure calculations with the (6,4) and (10,8) active spaces shown in Fig.\ \ref{fig:orbitals} are compared along the LIIC pathways, noting that the differences in geometries used to create the LIIC are negligible at either level of theory (see SI). The electronic structure calculations are found to be in quantitative and qualitative agreement across the LIIC. The validity of the  SA(4|4)-CASSCF(6,4)/ANO-R1(t) electronic structure calculations used in the simulations is further supported by benchmarks against other methods such as extended multi-state complete active space second-order perturbation theory, XMS-CASPT2, and multireference configuration interaction, MRCI;\cite{forsberg_multiconfiguration_1997, helgaker_molecular_2013} these results are also included in the SI.

\subsection{Dynamics}

The nonadiabatic dynamics of 1,2-dithiane is simulated with the trajectory surface hopping (TSH) method, using Tully's fewest switches algorithm,\cite{tully_molecular_1990} as implemented in the SHARC software package (version 2.1).\cite{oppel_sharc-mdsharc_2022,Mai2018WCMS,richter_sharc_2011}

From one set of $8000$ initial conditions sampled from a Wigner distribution at the S$_0$(min) structure, two ensembles are prepared: one containing only the first four lowest-energy singlet states ("singlet-only"), and another containing the same set of singlet states plus the first four lowest-energy triplet states ("triplet-inclusive").\cite{PhysRev.40.749,barbatti_effects_2016} We run 384 trajectories in each ensemble. An excitation window of 4.46 - 4.52 eV is used to excite the initial conditions for the dynamics; this window is based on a 290~nm UV pump-pulse with $30$~fs full-width-half-maximum (FWHM) pulse duration. We add a shift of 0.21 eV determined by fitting the simulated spectrum to an experimental gas-phase absorption spectrum.\cite{wreo20420,Bergson1962UltravioletAS} The choice of these specific parameters is motivated by the anticipation that such parameters may be used in future experiments.

The UV absorption spectrum is simulated using the nuclear ensemble approach.\cite{crespo-otero_spectrum_2012} This entails summing over the absorption spectra at all initial geometries, where the absorption spectrum at each initial geometry is the convolution of the line spectrum by a Gaussian function with FWHM $=0.2$ eV to represent the experimental energy resolution.

In both the singlet-only and triplet-inclusive ensembles, the trajectories are propagated for $1$ ps with a timestep of $0.5$~fs and a timestep of $0.02$~fs for the electronic integration. Dynamics are performed in the diagonal representation to allow for favourably localised couplings between electronic states, which in principle yields more accurate dynamics compared with those performed in the diabatic and adiabatic bases.\cite{Mai2018WCMS, granucci_surface_2012}  A local diabatisation scheme is used to calculate electronic gradients due to the greater efficiency compared with explicit calculations of nonadiabatic coupling matrix elements (NACMEs), particularly when the molecule adopts ring-opened geometries leading to dense manifolds of electronic states.\cite{granucci_critical_2007} The speed of the quantum chemistry calculations at each timestep is further improved by only calculating gradients when the difference in energy between electronic states is less than $0.5$~eV. This is justified on account of the inverse scaling of transition probabilities with respect to the energy gap between states; population transfer is less likely between more widely spaced states. The energy-difference based decoherence correction scheme developed by Granucci and Persico is implemented using the default recommended parameter of $0.1$~Hartree.\cite{granucci_including_2010} Following a surface hop, nuclear velocities are rescaled to adjust the kinetic energy to preserve the total energy. The total energy and any changes in energy between consecutive timesteps is closely monitored for all trajectories; energy conservation violations falling outside a $0.2$~eV threshold are flagged, and those trajectories are terminated at the timestep where the problem arises.

\subsection{X-ray scattering simulations}

For both sets of trajectory ensembles, ultrafast X-ray scattering (UXS) patterns are calculated using the independent atom model (IAM) approach.\cite{SimmermacherCh3Kasra,KirranderJCTC2016} The IAM assumes that the scattering patterns of a molecule, whether X-ray scattering or electron scattering, can be approximated as a coherent sum of tabulated isotropic scattering amplitudes of all constituent atoms.\cite{Minas2017} Any effects in the scattering signal which are caused by changes in electronic distribution resulting from bonding interactions within the molecule are thus not considered.\cite{Northey2014,Northey2016,Carrascosa2019total} In short, the rotationally averaged IAM elastic scattering, $I_\mathrm{el}(\tilde{{q}},\overline{\bm{R}})$, for a molecule containing $N$ atoms described by spatial coordinates $\overline{\bm{R}}=\{\bm{R}_1,\ldots,\bm{R}_N\}$ is given by\cite{SimmermacherCh3Kasra}

\begin{equation}
    I_\mathrm{el}(\tilde{{q}},\overline{\bm{R}}) = \sum_{A=1}^{N}\sum_{B=1}^{N}f_B^X(\tilde{q})f_A^X(\tilde{q})\frac{\sin(\tilde{q}R_{AB})}{\tilde{q}R_{AB}},
\label{eq:elastic}
\end{equation}

where $\tilde{q}=|\tilde{\bm{q}}|$ is the magnitude of the momentum transfer vector (within the Waller-Hartree approximation),\cite{waller_hartree} $f_i^X(\tilde{q})$ is the atomic form factor for atom $i$,\cite{affs} and $R_{AB}=|\bm{R}_{A}-\bm{R}_{B}|$ is the distance between atoms $A$ and $B$. The total X-ray scattering probability, $I_\mathrm{tot}(\tilde{{q}},\overline{\bm{R}})$, is then a sum of elastic and inelastic X-ray scattering components:
\begin{equation}
    I_\mathrm{tot}(\tilde{{q}},\overline{\bm{R}}) = I_\mathrm{el}(\tilde{{q}},\overline{\bm{R}}) + \sum_{A=1}^{N}S_A(\tilde{q}),
\label{eq:total}
\end{equation}
where the second term approximates the inelastic component of the total scattering and involves an incoherent sum of tabulated coherent scattering functions, $S_A(\tilde{q})$.\cite{affs} Thus, by using Eq.\ (\ref{eq:total}) to calculate the total scattering signal at every molecular geometry along a surface-hopping trajectory for all trajectories in an ensemble, the mean total scattering signals at each time -- that is, the mean simulated UXS signal -- can be calculated as
\begin{equation}
    I(\tilde{{q}},t) = \frac{1}{N_{\mathrm{trj}}} \sum_{i=1}^{N_\mathrm{trj}}I_\mathrm{tot}(\tilde{{q}},\overline{\bm{R}}_i(t)),
\label{eq:mean}
\end{equation}
with equal weight given to each trajectory $\overline{\bm{R}}_i(t)$. To evaluate how the UXS pattern varies across time it is convenient to plot the percent difference of the UXS signal, $\Delta\%I(\tilde{q},t)$, relative the signal at $t_0=0$:
\begin{equation} 
    \Delta\%I(\tilde{q},t) = 100\frac{I(\tilde{q},t) - I(\tilde{q},t_0)}{I(\tilde{q},t_0)}.
\label{eq:pdiff}
\end{equation}
We note that in the IAM, the inelastic component is invariant with respect to molecular geometry, meaning that only the elastic component affects the numerator in Eq.~(\ref{eq:pdiff}).\cite{Carrascosa2019total,Zotev2020}

\section{Results and discussion}

\subsection{Electronic structure of ground and excited states}
 
 The LIIC potential energy curves, the electronic state characters and the corresponding CI coefficients at the S$_0$(min) geometry, as well as the molecular structures at the S$_0$(min), S$_1$(min), and MECI(S$_0$,S$_1$) geometries, are all shown in Fig.\ \ref{fig:LIIC}. The ground state minimum, \ce{S_0}(min), adopts a chair-like conformation in a similar manner to cyclohexane. Previous literature suggests that there also exists a local minimum twist-boat structure at approximately $10$--$20$~kJ mol$^{-1}$ above the chair minimum, which may be accessed {\it{via}} a half-chair transition state lying approximately $50$~kJ mol$^{-1}$ above the ground state minimum.\cite{doi:10.1080/10426500490473537,dd60e77fdf444c4eb252f6c4616b994c} In the present calculations, using SA(4|4)-CASSCF(6,4)/ANO-R1(t), the twist-boat energy was found to lie $23$ kJ mol$^{-1}$ above the ground state. Assuming a two-state Boltzmann distribution, and accounting for the vibrational zero point energy of each conformer, only $0.5$\% of the molecules are expected to adopt the twist-boat geometry at $298$~K. Thus, only initial conditions from the chair-like minimum are considered in the following, with further details on the twist-boat conformation provided in the SI.
 
 In both the \ce{S_1}(min) and the MECI(\ce{S_0},\ce{S_1}) geometries, the \ce{S-S} bond is broken, and the two sulfur atoms have radical character. The \ce{S-S} distance for the \ce{S_0}(min) geometry is $2.13$~\AA, while for the \ce{S_1}(min) and MECI(\ce{S_0},\ce{S_1}) geometries, the \ce{S-S} distances are $3.64$~\AA\ and $3.91$~\AA, respectively. The MECI geometry is thus quite similar to the \ce{S_1} minimum, with the root mean square deviation (RMSD) between the two Kabsch-rotated structures equal to $0.0914$~\AA. In contrast, the RMSD between \ce{S_1}(min) and \ce{S_0}(min) is $0.3529$~\AA.

The LIIC pathway shown in Fig.\ \ref{fig:LIIC} is constructed by interpolation from S$_{0}$(min) to S$_{1}$(min), and then onwards to the MECI(\ce{S_0},\ce{S_1}), \textit{i.e.}\ S$_{0}$(min)$\rightarrow$S$_{1}$(min)$\rightarrow$MECI(\ce{S_0},\ce{S_1}). The main change in molecular geometry along the LIIC pathway is a progressive stretching of the \ce{S-S} bond; there are only minor changes in other internal coordinates, such as a small rotation in the carbon backbone dihedral angle. Looking at Fig.\ \ref{fig:LIIC}(a), it is evident that the excited \ce{S_1}--\ce{S_3} and \ce{T_1}--\ce{T_4} states are dissociative in the Franck-Condon (FC) region near S$_0$(min). As the \ce{S-S} bond stretches, the potential energies of the excited states decrease smoothly, thus providing a barrierless path to the \ce{S_1} minimum. From left to right along the LIIC pathway, the \ce{S_0} energy increases. As the \ce{S-S} distance increases, all potential energy curves considered approach similar asymptotic energies corresponding to a broken \ce{S-S} bond, with no or very little overlap between the orbitals on the two sulfur atoms. 

The dominant electronic state characters for the S$_0$-S$_3$ and T$_1$-T$_4$ states are listed alongside their configuration interaction (CI) coefficients in Table \ref{fig:LIIC}(b).\footnote{The double-excitation character of \ce{S_3} and \ce{T_4} comprises two separate double excitations from the two sulfur lone-pair orbitals - n$^\mathrm{u}_\mathrm{S}$ and n$^\mathrm{g}_\mathrm{S}$ - displayed in Fig.\ \ref{fig:orbitals}.} The \ce{S_0} state in the FC region is defined predominantly by two electrons each in the $\mathrm{\sigma_{S-S}}$, $\mathrm{n^g_{S}}$, and $\mathrm{n^u_{S}}$ valence orbitals. Given that all excited states considered here populate the \ce{S-S} $\sigma^*$ orbital, this qualitatively explains the repulsive nature of these states with respect to the \ce{S-S} stretching coordinate, as seen in Fig.\ \ref{fig:LIIC}. By comparing the state characters in the FC region between the $(6,4)$ and $(10,8)$ active spaces, we see that the \ce{C-S} orbitals are unlikely to influence the state characters at energetically accessible geometries considering the excitation window in the simulations, further supporting the use of the smaller $(6,4)$ active space in the simulations.

Finally, all SOCMEs were calculated between the four lowest-energy singlet and four lowest-energy triplet states at geometries along the \ce{S_0}--\ce{S_1} LIIC pathway. Across the LIIC pathway, strong spin-orbit coupling with SOCMEs greater than $100$~cm$^{-1}$ ($0.01$~eV) is present between both singlet-triplet and triplet-triplet state pairs, as may be anticipated from the $Z^4$ scaling of spin-orbit coupling in an atomistic picture. Such strong SOC effects should be expected to facilitate efficient population redistribution among the triplet states from the singlet manifold. This supports and justifies the need to investigate the effect of including the triplet states in the simulations.

\subsection{UV absorption spectrum}

\begin{figure}[t]
\centering
\includegraphics[width=0.49\textwidth]{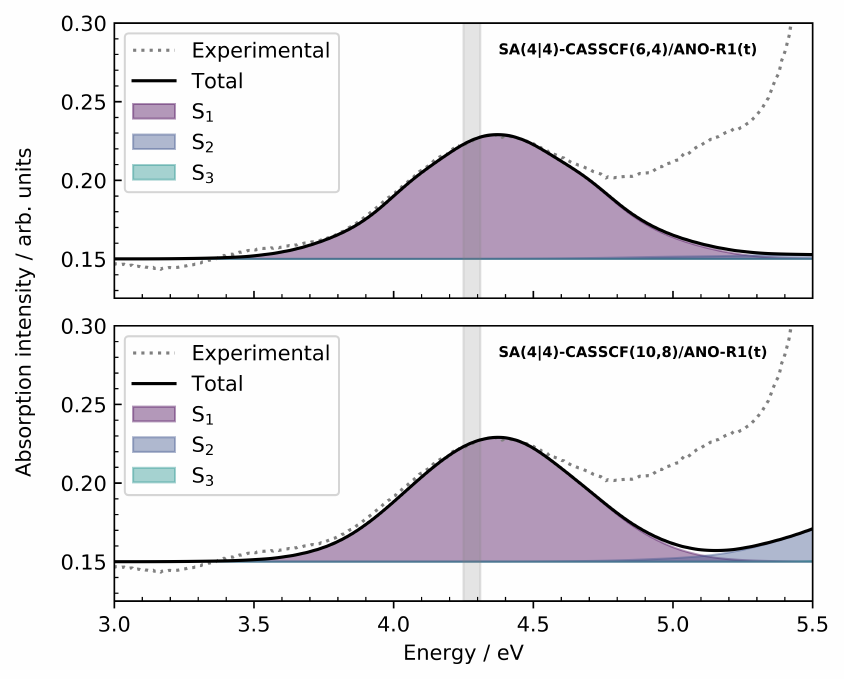}
 \caption{Experimental\cite{wreo20420,Bergson1962UltravioletAS} (dotted line) and computational UV absorption spectra (shifted by -$0.21$ eV) for 1,2-dithiane calculated at the SA(4$\vert$4)-CASSCF(6,4)/ANO-R1(t) (top) and the SA(4$\vert$4)-CASSCF(10,8)/ANO-R1(t) (bottom) levels of theory, respectively. The proposed excitation window, $4.25$--$4.31$~eV is indicated by the shaded light-grey region on both plots. For a discussion of the strong absorption at energies $>4.5$~eV in the experimental spectrum, see the text.}
 \label{fig:spectrum}
\end{figure}

The UV absorption spectrum for 1,2-dithiane, calculated at the SA(4$\vert$4)-CASSCF(6,4)/ANO-R1(t) level of theory, is shown in the top panel of Fig.\ \ref{fig:spectrum}. The absorption spectrum calculated with the $(10,8)$ active space is presented in the bottom panel to aid with the assignment of the observed features in the experimental absorption spectrum.\cite{wreo20420,Bergson1962UltravioletAS} Both the experimental and shifted computational spectra show a broad peak with a maximum centred at approximately $4.37$~eV, which corresponds to direct excitation into the \ce{S_1} state. For both active spaces, much of the intensity of this peak can be ascribed to a $\sigma^*_{\mathrm{S-S}}\leftarrow $ n transition. This peak energy is in reasonable agreement with the vertical excitation energy from S$_0$ to S$_1$ ($\Delta E_{\mathrm{S}_1,\mathrm{S}_0}=4.50$~eV) as shown in Fig.\ \ref{fig:LIIC}.

\begin{figure*}[h]
\centering
{\includegraphics[width=0.49\textwidth]{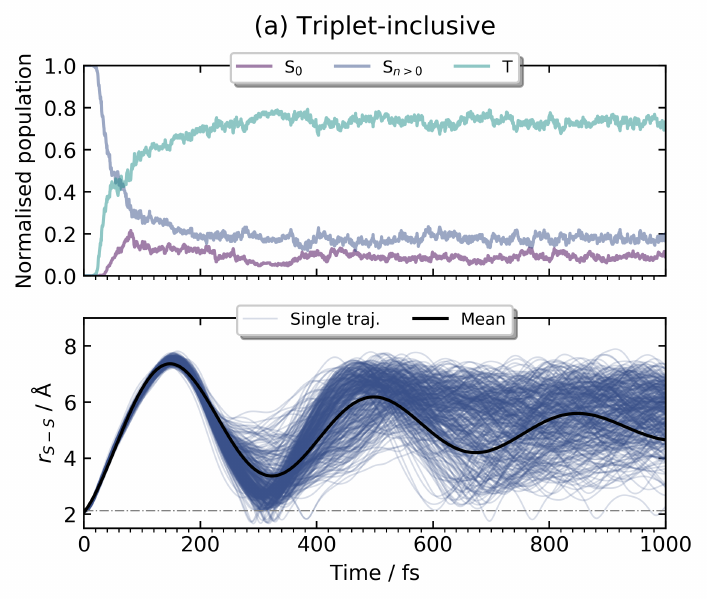}}
{\includegraphics[width=0.49\textwidth]{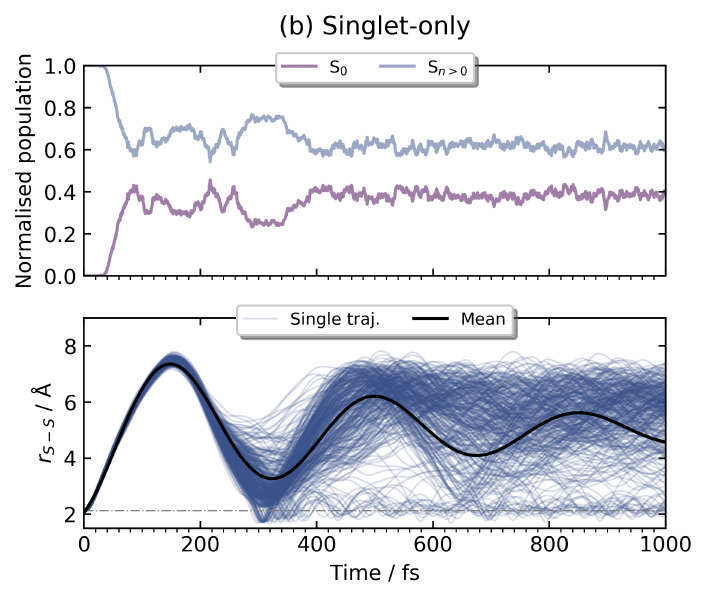}}
\caption{Populations and \ce{S-S} distances as a function of time (fs) for the triplet-inclusive (left column) and the singlet-only (right column) simulations. (Top row) The classical molecular Coulomb Hamiltonian (MCH) state representation populations. The left panel shows the ground state population \ce{S_0}, the net singlet excited-state population \ce{S_n} (summing the populations for all singlet states with $n>0$), and the net triplet population \ce{T} (summing over all triplet states). The right panel is the same, but excluding the net triplet population. (Bottom row) The \ce{S-S} bond length, $r_{\mathrm{S-S}}$, for each trajectory with the mean bond length fitted to an exponentially damped sine function and shown as a thick black curve, analogously to the fitting in Rankine \textit{et al}.\cite{rankine_theoretical_2016} For reference, the grey dot-dashed line indicates the \ce{S-S} distance at the \ce{S_0} equilibrium geometry, 2.13~\AA.}
\label{fig:pop_rSS}
\end{figure*}

Noticeable deviations from the experimental gas-phase spectrum occur for both calculated absorption spectra at energies greater than $4.5$~eV. A very strong absorption peak is seen in the experimental spectrum at energies exceeding $5.5$~eV; a shoulder feature at approximately $5.2$~eV is also seen in this spectrum.  Excitation into the S$_2$ state, corresponding for the most part to a $\sigma^*_{\mathrm{S-S}}\leftarrow $ n transition, only marginally accounts for absorption intensity for energies exceeding $4.5$~eV when the $(10,8)$ active space is used. This strongly suggests that excitation into states which are not included in the $(6,4)$ and $(10,8)$ active spaces, for example Rydberg states, must be responsible for the majority of the absorption intensity in this region.\cite{KIMBER202455} Nonetheless, the excitation window used for the dynamics is well separated from this region, supporting the use of the smaller active space for the simulations.

\subsection{Simulations}

In total, $384$ trajectories are propagated for both the triplet-inclusive and singlet-only trajectory ensembles, with all dynamics initiated on the \ce{S_1} state. For the triplet-inclusive ensemble, $371$ trajectories (96.6\%) were propagated successfully, while for the singlet-only ensemble, $346$ trajectories (90.1\%) were propagated successfully. A diagnostic analysis of the prematurely terminated trajectories is provided in the SI.

The classical populations, calculated within the molecular Coulomb Hamiltonian (MCH) representation of states,\cite{Mai2018WCMS} as well the changes in \ce{S-S} bond distance, $r_{\mathrm{S-S}}$, as a function of time are shown in Fig.\ \ref{fig:pop_rSS}. Both simulations exhibit a rapid decrease in the excited singlet-state population, S$_{n>0}$, dominated by loss of population from the \ce{S_1} state in the period $20 < t < 90$~fs. This is mirrored by a commensurate increase in the \ce{S_0} population, in particular for the singlet-only simulation. In the simulation that also includes the triplet states, a rapid increase in the net triplet population dominates over the increase in \ce{S_0} population.

The mean times at which the first surface hopping events occur are $27.5\pm 6.4$~fs and $32.9\pm 10.4$~fs for the triplet-inclusive and singlet-only simulations, respectively. These values agree well with the timescales for depopulation of the initially excited \ce{S_1} state reported in previous studies.\cite{rankine_theoretical_2016,cao_disulfide_2019,ibele_comparing_2021} The ultrafast nonadiabatic transfer out of \ce{S_1} is associated with stretching and subsequently cleaving the \ce{S-S} bond, as seen in the plots in Fig.\ \ref{fig:pop_rSS}, which is also commensurate with the dissociative LIIC pathway shown in Fig.\ \ref{fig:LIIC}. The ground and excited states rapidly come together as the \ce{S-S} bond stretches towards the MECI(\ce{S_0},\ce{S_1}) facilitating effective population transfer. The \ce{S-S} bond length is already greater than 3~\AA~at $t=25$~fs, with most electronic states within a narrow $0.1$~eV energy band at this point. The more rapid decay of the \ce{S_1} population in the triplet-inclusive case correlates well with the overall higher density of states in the triplet-inclusive case and thus the availability of more decay channels. Strikingly, the net triplet population, \ce{T_n} rises \textit{earlier} than the S$_0$ population in the triplet-inclusive simulations, indicating that the proximity of singlet and triplet states already at short \ce{S-S} distances leads to rapid and efficient ISC even before the onset of IC to the ground state. We also observe a moderate rise in the S$_2$ population -- before any significant rise in the populations of any other electronic state -- at approximately $t=20$~fs in both trajectory ensembles; see SI for the variation in the populations of individual electronic states across the total trajectory time. This suggests a dominant S$_2 \leftarrow$ S$_1$ pathway before population redistribution to other electronic states --- especially triplet states --- occurring immediately after. Reassuringly, Cao and Chen also make this observation in their investigation into the early-time population dynamics of 1,2-dithiane.\cite{cao_disulfide_2019}

\begin{figure*}[h]
\centering
{\includegraphics[width=\textwidth]{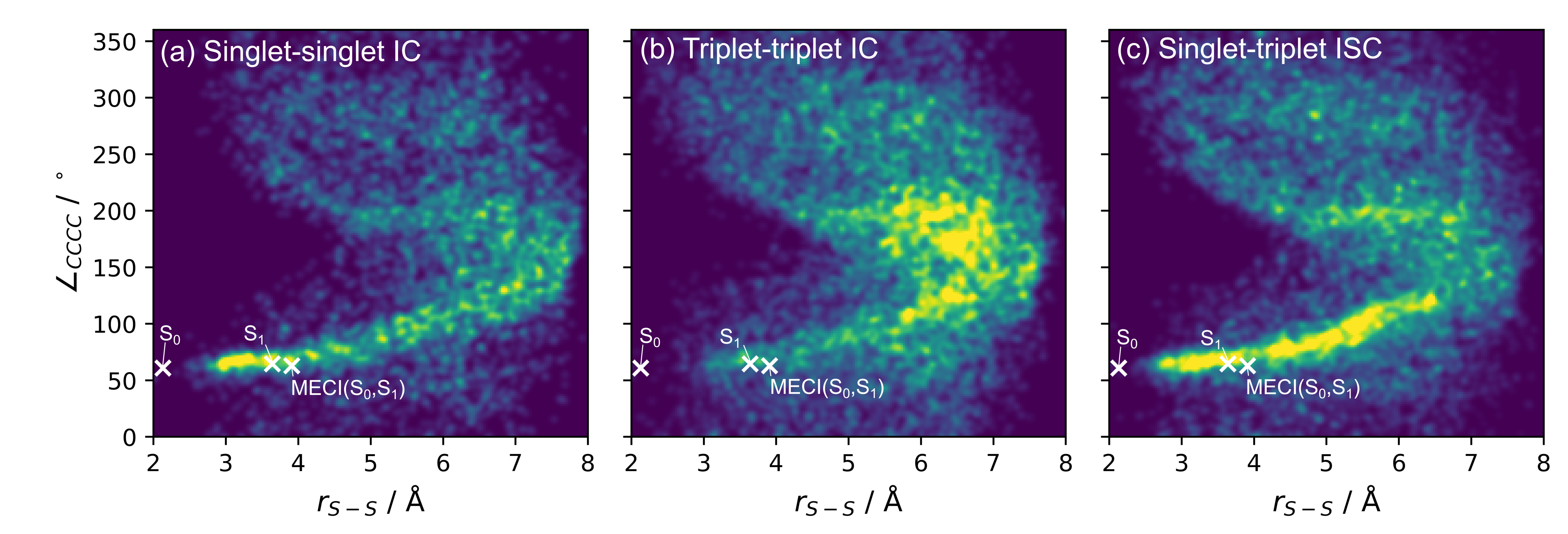}}
\caption{Density heatmaps of crossing geometries --- with respect to the disulfide distance, $r_{\mathrm{S-S}}$, and the carbon backbone dihedral angle, $\angle_{CCCC}$ --- for all surface hopping events occurring in the triplet-inclusive trajectory ensemble. Surface hops are partitioned into subsets involving (a) singlet-singlet IC, (b) triplet-triplet IC, and (c) singlet-triplet ISC processes. The S$_0$(min), S$_1$(min), and MECI(S$_0$,S$_1$) geometries are indicated on each heatmap for reference. Details of additional electronic state minima and MECIs which were located --- not shown here for clarity --- can be found in the SI.}
\label{fig:heatmap}
\end{figure*}

Both simulations show a similar trend with respect to the \ce{S-S} distance, $r_{\mathrm{S-S}}$. The initial concerted ring-opening progresses from $t=0$ to form linear dithiyl structures, reaching maximum values of $r_{\mathrm{S-S}}$ at approximately $t=150$~fs before the thiyl termini return to close proximity at $t\approx320$~fs. The ring-closing dynamics during $150<t<320$~fs is associated with a steady \ce{S_0} population, owing to the large energy gap between \ce{S_0} and all excited states as $r_{\mathrm{S-S}}$ approaches the equilibrium geometry \ce{S_0} bond length. The trajectories on \ce{S_0} during the ring-closing phase thus tend to remain on \ce{S_0} at least until the ring begins to open again. From approximately $320$~fs, nuclear motion starts to significantly decohere as the vast majority of trajectories in both ensembles undergo a second ring-opening. Subsequently, most trajectories in both simulations remain in ring-open geometries for the remainder of the simulations. Fitting the time dependence of $r_{\mathrm{S-S}}$ to damped sine functions yields oscillatory time periods of $350.2\pm1.1$~fs and $351.1\pm1.2$~fs for the triplet-inclusive and singlet-only simulations, respectively. Remarkably, both time periods are within statistical error of each other and agree within error with the value of $349.6$~fs reported by Rankine et al.\cite{rankine_theoretical_2016} Such quantitative agreement suggests that the nuclear dynamics is quite robust with respect to the number of electronic states considered and the electronic structure method used, with the main point for the latter being to reproduce the dissociative nature of the excited states.

The loss of concerted nuclear motions from approximately $400$~fs correlates well with the equilibration of state populations seen in both simulations. Despite the fact that individual trajectories may still undergo oscillatory motions after the initial ring-opening and closing, the sulfur termini generally remain several \AA ngstr\"{o}ms apart for the majority of trajectories at these later times in both simulations. Thus, trajectories are almost always found in a region where the electronic state density is high --- and thus also in a region where IC and ISC is efficient --- resulting in a high probability of surface hopping. Indeed, the ubiquity of surface hops in this time range facilitates population equilibration of singlet and triplet states towards an almost statistical limit. At $t=1$~ps, the total triplet population is $70.9$\%. Given the 12 triplet states and 4 singlet states included in the simulations, a statistical 75\% triplet yield could be anticipated at longer times beyond $1$~ps. 

In addition, population transfer between electronic states is not localised to specific geometries, \textit{i.e.}\ IC and ISC are not dominated by any specific conical intersections or minimum energy crossing points in this system. This is illustrated in Fig.\ \ref{fig:heatmap}, which shows the distribution of surface-hopping events in the triplet-inclusive ensemble for all IC and ISC transitions across the total trajectory time. The distribution of surface hops across a large region of molecular geometries in both simulations, for both IC and ISC, is a direct consequence of the comparatively flat potential energy landscape outside the ground state equilibrium well.

Returning to Fig.\ \ref{fig:pop_rSS}, both ensembles show very similar trends with respect to the variation of $r_{\mathrm{S-S}}$ across the trajectory time. However, several trajectories exhibit oscillations in $r_{\mathrm{S-S}}$ about the \ce{S_0} equilibrium disulfide bond length from approximately $300$~fs in the singlet-only ensemble; these features are not observed in the triplet-inclusive ensemble. At $300$~fs, 51/346 (14.7\%) successful trajectories in the singlet-only ensemble propagate in the ground state potential well, whereas only 15/371 (4.0\%) in the triplet-inclusive ensemble. Those in the former set are wholly responsible for the trajectories which show $r_{\mathrm{S-S}}$ becoming temporarily compressed below the \ce{S_0} equilibrium value (2.13 \AA) at approximately $300$~fs in the lower subplot in Fig.\ \ref{fig:pop_rSS}(b). 

The presence of the triplet states effectively prevents trajectories from entering the \ce{S_0} potential well. All trajectories that enter the \ce{S_0} potential well at $300$~fs in the triplet-inclusive ensemble eventually exit and return to ring-opened conformations. This reflection out of the ground state potential energy well results from the repulsive force between sulfur atoms as $r_{\mathrm{S-S}}$ is compressed beyond the \ce{S_0} equilibrium value. For the singlet-only simulations, out of the $51$ trajectories in the \ce{S_0} potential well at $t=300$~fs, $8$ remain on the \ce{S_0} potential energy well for the rest of the simulation, giving rise to the oscillations in $r_{\mathrm{S-S}}$ around the equilibrium \ce{S_0} distance in Fig.\ \ref{fig:pop_rSS}(b); the other $43$ trajectories return to the ring-open coupling region. It is possible that permanent \ce{S-S} recombination may have been observed in the triplet-inclusive ensemble had an even greater number of trajectories been propagated. Nonetheless, it is clear that the recombination pathway is  significantly disfavoured when ISC processes are accounted for when the total number of trajectories is fixed between both ensembles.

Trajectories that are not on \ce{S_0} are deflected by the dissociative nature of the excited states. This repulsive character of the excited state potentials with respect to $r_{\mathrm{S-S}}$ is the main factor behind the oscillatory {\it{Newton's cradle}}-type nature of most trajectories. Oscillations about the \ce{S_0} equilibrium bond length in Fig.\ \ref{fig:pop_rSS}(b) --- which have a smaller amplitude and shorter period than oscillations resulting from the {\it{Newton's cradle}}-type behaviour --- correspond to the eight trajectories which remain indefinitely on the ground state in this ensemble. The absence of such oscillations in the triplet-inclusive ensemble --- with the exception of one trajectory which remains in the potential well from $870$~fs onwards --- is a consequence of the reduced probability of being in the \ce{S_0} state at any given time. This can be seen by comparing the trends in \ce{S_0} populations in Fig.\ \ref{fig:pop_rSS}. Notably, the proportion of trajectories remaining in the \ce{S_0} potential well after the first ring-closing event is very low (2.3\%), even for the singlet-only ensemble.

\begin{figure}
\centering
\includegraphics[width=0.45\textwidth]{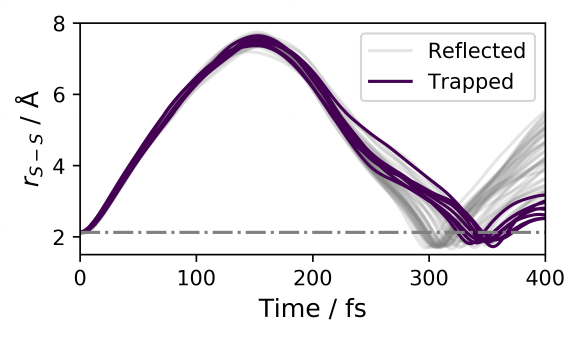}
\caption{The variation in $r_\mathrm{{S-S}}$ over the first 400 fs of dynamics simulations for all trajectories which descend into the ground state potential well upon the first ring-closing event for the singlet-only ensemble. Trajectories which reflect out of the potential well are shown in grey, whilst those which remain trapped on \ce{S_0} for the rest of the simulation time are shown in purple.}
\label{fig:rSS_trapped_ref}
\end{figure}

\begin{figure*}[h]
\centering
{\includegraphics[width=0.95\textwidth]{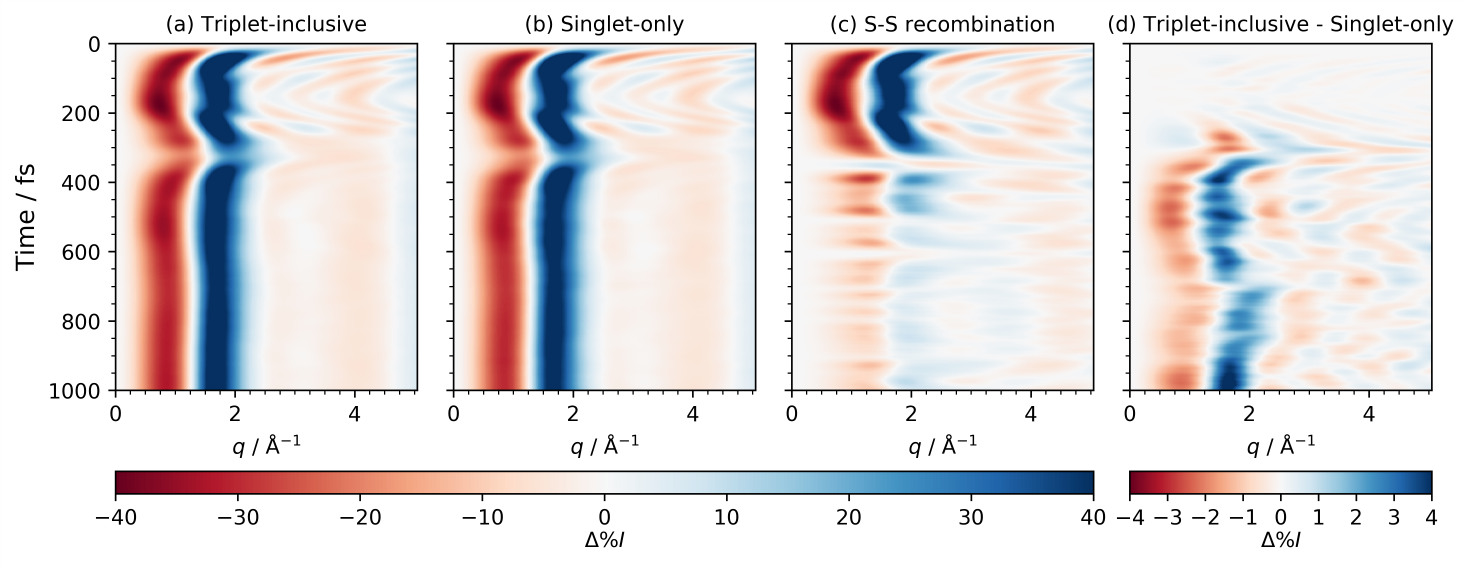}}
\caption{Rotationally averaged total X-ray scattering patterns for the simulations as a function of time. The scattering signal is expressed as a percent difference signal according to Eq.\ \ref{eq:pdiff}, and is shown for the (a) triplet-inclusive and (b) singlet-only simulations, as well as (c) the subset of trajectories from the singlet-only simulation where \ce{S-S} radical recombination occurs following the first ring-closing event at approximately $t=350$~fs. The final panel (d) shows the difference between the signal shown in panels (a) and (b).}
\label{fig:iam}
\end{figure*}

Considering the singlet-only trajectories which enter the ground state potential well, the variation in $r_{\mathrm{S-S}}$ over the first $400$~fs of simulations --- separated into a subset which reflects out of the well and a subset which remains trapped for the duration of the simulation --- is shown in Fig.\ \ref{fig:rSS_trapped_ref}. Whilst both trajectory subsets display coherent nuclear dynamics with respect to $r_{\mathrm{S-S}}$ over the first 200 fs, they deviate in how individual trajectories navigate through the potential well conformational landscape. Trajectories which ultimately return to the strong nonadiabatic coupling region generally exhibit turning points in $r_{\mathrm{S-S}}$ approximately 50~fs earlier than those which remain on \ce{S_0}, thereby indicating a steeper descent into the well with respect to electronic state gradients. This causes more rapid $r_{\mathrm{S-S}}$ bond compression as $r_{\mathrm{S-S}}$ decreases below the ground-state equilibrium value, thus resulting in a more forceful collision of sulfur atoms as the linear momentum along the \ce{S-S} stretching coordinate is approximately conserved in a shorter collision time frame. Conversely, trajectories which remain trapped on \ce{S_0} display a more gentle descent into the potential valley. This gives a greater period of time during which energy can be redistributed to other internal degrees of freedom, thereby resulting in less available kinetic energy in the appropriate internal degrees of freedom for escaping the ground state. 

Differential rates of vibrational relaxation and internal energy redistribution between individual trajectories also rationalises both the damped nature of subsequent oscillations in $r_{\mathrm{S-S}}$ for both individual trajectories and the fitted mean, and also the nuclear decoherence across both ensembles following the first ring-opening and ring-closing events. The latter observation results in ergodic ensemble dynamics over increasingly longer time periods as the excited state conformational landscape is explored more thoroughly.

\subsection{X-ray scattering signals}

Whilst slight differences in nuclear dynamics -- namely the minor \ce{S-S} recombination pathway in the singlet-only ensemble -- are evident in the simulations, it is interesting to ascertain whether this difference may give rise to distinct and detectable signals in an experiment. Provided that there is sufficient spatial and temporal resolution, where the latter is ideally on the order of tens of femtoseconds or less, UXS experiments can be particularly sensitive to changes in molecular structure on ultrafast timescales.\cite{BudarzJPB2016} Using the IAM to calculate UXS patterns is also straightforward and computationally inexpensive in that it only requires a set of molecular geometries and tabulated atomic form factors\cite{affs}; no prior knowledge of the electronic structure of the system is required. Despite the approximations inherent in the IAM, it is nonetheless a useful first-order approximation for calculating UXS patterns, particularly in light of the moderate molecular size of 1,2-dithiane and the relatively large number of intermediate timesteps involved in a single trajectory which would render more complex UXS simulations using {\it{ab initio}} electronic structure theory computationally expensive.\cite{SimmermacherCh3Kasra}

The mean percent difference UXS signals for both simulations, the difference in the UXS signals between both simulations, and the signal arising from the subset of trajectories which display permanent \ce{S-S} recombination in the singlet-only simulations, are all shown in Fig.\ \ref{fig:iam}. Strikingly, the triplet-inclusive and singlet-only trajectory ensembles yield very similar UXS signals owing to almost identical mean nuclear dynamics. However, some differences persist, in particular for times $t>350$~fs, as can be seen in Fig.\ \ref{fig:iam}(d). Whilst these differences are roughly an order of magnitude smaller than the overall scattering signal, they could in principle be detected, particularly considering the excellent signal-to-noise anticipated for new high-repetition rate experiments at X-ray free-electron laser (XFEL) facilities such as the upgraded Linac Coherent Light Source (LCLS-II).\cite{Carrascosa2022} 

Further to this, the UXS signal arising from trajectories involving \ce{S-S} recombination, shown in Fig.\ \ref{fig:iam}(c), is qualitatively similar to the total singlet-only signal in panel (b) for times $t<350$~fs. This is due to the concerted Newton's cradle ring-opening/closing nuclear dynamics which is dominant during these initial stages of the reaction. However, Fig.\ \ref{fig:iam}(c) \textit{does} deviate from the signals in panels (a) and (b) for times $t>350$~fs. This is most clearly seen by the depletion in Fig.\ \ref{fig:iam}(c) of the strongly negative and strongly positive features centred at $q=0.8$~\AA$^{-1}$ and $q=1.7$~\AA$^{-1}$, respectively, which can be observed in panels (a) and (b) from 350~fs onwards. The recombination trajectories display weakly positive UXS bands for $2.5$~\AA$^{-1}<q<4.0$~\AA$^{-1}$, whereas the ensemble averaged UXS signals in Figs.\ \ref{fig:iam}(a-b) are weakly negative in this region. It remains to be seen whether the scattering patterns in Fig.\ \ref{fig:iam}(a), which derive from the more physically realistic simulation that includes the triplet states, is in better agreement with UXS signals measured in future experiments, and if those experiments will require that the scattering is calculated using \textit{ab-initio} methods.\cite{Carrascosa2019total}

We finally note that this study may somewhat underestimate the role of recombination in the singlet-only simulation as compared to the triplet-inclusive. The reason for this is that the electronic structure stability issues which prematurely terminate some trajectories occur when $r_{\mathrm{S-S}}$ is compressed and the molecule propagates on S$_0$ state. This issue disproportionally affects the singlet-only simulations, as explained in further detail in the SI.

\section{Conclusions}

Simulations of the photoexcited dynamics in 1,2-dithiane on ultrafast timescales have demonstrated the characteristic oscillatory \textit{"Newton's cradle"} molecular behaviour, which is shown to be present irrespective if triplet states are included or not. In both simulations, there is rapid population transfer and redistribution which commences approximately 30~fs after excitation. The rapid population transfer arises from the concerted ring-opening, which brings ground and excited electronic states into a narrow energy band as the \ce{S-S} bond is elongated beyond dissociation. While both singlet and singlet-triplet simulations exhibit almost identical mean variation of the $r_{\mathrm{S-S}}$ distance, the inclusion of the triplet states noticeably deactivates the dithiyl radical recombination pathway observed from $\sim$300~fs onwards in the singlet-only simulations. This is a consequence of fewer trajectories descending into the ground-state S$_0$ potential well in the triplet-inclusive simulation, which can be seen as a statistical effect, with population distributed to the inactive triplet states. However, owing to the scarcity of \ce{S-S} recombination events even in the singlet-only simulation, the effect on the predicted UXS signal is comparatively small, at least at the level of the IAM approximation. The differences in UXS signal between the two simulations nonetheless may still be experimentally resolvable. It is also worth noting that other experimental techniques, such as time-resolved photoelectron imaging,\cite{Townsend2021} may be better equipped to differentiate the populations on the singlet and triplet states. If anything, this strengthens the broad conclusion of this paper that the triplet states should be accounted for when considering the dynamics of this system and related molecular disulfides. 

In terms of structural dynamics, the nuclear decoherence and equilibration of excited states after $400$~fs are shared by both simulations and reflect the tendency for redistribution of kinetic energy and linear momentum along the \ce{S-S} bond axis to other internal degrees of freedom at longer times. This behaviour is clearly quite robust and predicted irrespective of what electronic states are included in the simulations. Future ultrafast imaging experiments on 1,2-dithiane proposed at the European XFEL facility will hopefully eventually verify --- or disprove --- our predictions for the nuclear dynamics. 

The hypothesis that the triplet states deactivate the dithiyl recombination pathway in 1,2-dithiane may at first glance seem at odds with the experimental observation that disulfide linkages in proteins are particularly resilient to UV photolysis. However, there are differences between the currently examined small molecule gas-phase reaction and the \ce{S-S} moiety in proteins. In proteins, polar or sterically bulky amino acid side chains in the vicinity of disulfide bridges, including those which may participate in hydrogen bonding, limit the conformational space available to the disulfide moiety upon S-S bond homolysis. This is likely to promote a more efficient and accessible radical recombination so as to retain the three-dimensional structure of the protein, and consequently its biological function, even on sub-picosecond timescales. By taking such effects into consideration in future studies, and by investigating how variations in disulfide molecular geometry and the carbon linkage affect the underlying electronic structure and dynamics,\cite{mcghee_2024} a more holistic understanding of the impressive stability of disulfide bonds in proteins can hopefully be achieved.

\section*{Author contributions}
{\bf{James Merrick:}} Conceptualisation (supporting); data curation (lead); formal analysis (lead); investigation (lead); methodology (lead); software (lead); validation (lead); visualisation (lead); writing -- original draft (lead); writing -- review and editing (lead). {\bf{Lewis Hutton:}} Software (supporting); writing -- review and editing (supporting). {\bf{Joseph C. Cooper:}} Conceptualisation (supporting); writing -- review and editing (supporting). {\bf{Claire Vallance:}} Conceptualisation (supporting); funding acquisition (equal); project administration (equal); supervision (equal); writing -- review and editing (supporting). {\bf{Adam Kirrander:}} Conceptualisation (lead); funding acquisition (equal); project administration (equal); resources (lead); supervision (equal); writing -- review and editing (supporting).

\section*{Conflicts of interest}
There are no conflicts to declare.

\section*{Data availability}
Data supporting this work, including geometry optimisation and electronic structure benchmarking calculations in addition to auxiliary dynamics analyses, can be found in the Supplementary Information (SI).

\section*{Acknowledgements}
JM would like to thank Mats Simmermacher for helpful discussions and support in running the X-ray scattering calculations, and acknowledges a doctoral studentship from the Engineering and Physical Sciences Research Council (EPSRC UK) and the \textit{Carolyn and Franco Gianturco Theoretical Chemistry Scholarship} from Linacre College at the University of Oxford. LH and AK acknowledge the support of the Leverhulme Trust grant RPG-2020-208, and JCC and AK acknowledge EPSRC grant EP/X026698/1. CV acknowledges funding from EPSRC Programme Grants EP/V026690/1 and EP/T021675/1. Finally, AK acknowledges funding from EPSRC grants EP/V006819/2, EP/V049240/2, and EP/X026973/1, as well as the U.S.\ Department of Energy, Office of Science, Basic Energy Sciences, under award number DE-SC0020276.



\balance


\bibliography{ref} 
\bibliographystyle{rsc} 

\end{document}